\documentclass[aps,prl,twocolumn,reprint,letterpaper]{revtex4-2}

\usepackage{graphicx}     
\usepackage{subfigure}
\usepackage{hyperref}
\usepackage{bm}
\usepackage{ae}
\usepackage{aecompl}
\usepackage{amsmath}
\usepackage{amssymb}
\usepackage{epsfig}
\usepackage{psfrag}
\usepackage{subfigure}
\usepackage{textcomp}

\newcommand{\Eq}[1]{Eq.~(\ref{eq:#1})}
 
\newcommand{\hh}{h}                
\newcommand{\x}{{\bf x }}                
\newcommand{\LL}{\xi_{\parallel}}                

\usepackage{color}

\newcommand{\rechange}[1]{{{#1}}}
\begin{document}
	
	\preprint{APS/123-QED}

\title{Surface tension of bulky colloids, capillarity under gravity,
  \\  and the microscopic origin of the Kardar-Parisi-Zhang equation}


\author{Luis G. MacDowell}
\affiliation{Departamento de Qu\'{\i}mica-F\'{\i}sica, 
   Facultad de Ciencias Qu\'{\i}micas, Universidad Complutense de Madrid, 28040
   Madrid, Spain.
}

\email[]{lgmac@quim.ucm.es}




\begin{abstract}
  Experimental measurements of the surface tension of colloidal interfaces have long been in conflict with computer simulations. In this work we show that the surface tension of colloids as measured by surface fluctuations picks up a gravity dependent contribution which removes the discrepancy. The presence of this term puts a strong constraint on the structure of the interface which allows one to identify {corrections} to the fundamental equation of equilibrium capillarity and deduce bottom-up the microscopic origin of a growth model with close relation to the Kardar-Parisi-Zhang equation.
\end{abstract}


\maketitle


A student can easily measure the surface tension of water 
using a modest equipment such as a Nouy ring availabe in 
undergraduate labs. As the Nouy ring is lifted gently with a spring against
surface tension {\em and gravity}, an equilibrium is established 
which reproducibly yields $\gamma=72$~mNm$^{-1}$ at room temperature.
But is this result affected by earth's gravity?

Admittedly, this question looks odd at first thought. But an important
consequence of renormalization theory is that
interfaces must exhibit small perpendicular fluctuations of the local 
interfacial position which are damped by  gravity.\cite{buff65,zittartz67,jasnow84} 
Whereas small in amplitude,
the interfacial fluctuations remain correlated over extremely
large distances, corresponding to the parallel correlation length
or capillary distance, $\LL^2 = \gamma / \Delta \rho\, G$ as set by the
gravitational acceleration, $G$  (with  $\Delta\rho$
the density difference between the bulk phases).
However, this widely accepted result poses a serious problem in the
limit of strong fields. Indeed, as $G$ becomes large, it predicts a vanishing
parallel correlation length, while on expects that $\LL$ should have
a lower bound that is dictated by the bulk molecular correlation
length of the fluid.\cite{rowlinson82b}
Interestingly, the correct large and small limits of $\LL$ may be
enforced heuristically by assuming  a gravity dependent surface tension:
\begin{equation}\label{eq:gravity}
\gamma(G) = \gamma_0 + \xi^2 \Delta \rho\, G,
\end{equation} 
with $\gamma_0$ the surface tension in absence of an external
field, and $\xi$, a measure of the bulk correlation length.\cite{macdowell13}

Unexpected as this may be, the result of \Eq{gravity}  is difficult to rule out 
for a molecular fluid well away from the critical point.
In view of the smallness of the bulk correlation length, which rarely
is larger than a few molecular diameters, the gravity dependent
term may be estimated on the order $10^{-11}$mN$m^{-1}$ for water at
room temperature, an unmeasurable correction that is a trillion times 
smaller than water's actual surface tension.

However, statistical mechanics has been borrowing experimental
results from colloidal science for more than 30
years.\cite{pusey86,vanblaaderen95,aarts04} Indeed, 
bulky colloids of micrometer size are regularly exploited to test predictions
for simple models of atomic interactions, as their size allows direct
optical observation. 

A paradigmatic example is the `hard sphere' colloid,
which exhibits a freezing transition and  packing correlations that
are in quantitative agreement with hard sphere results obtained from computer
simulations.\cite{pusey86,vanblaaderen95}
By use of confocal 
microscopy, the interface that is formed can be observed and
analized.\cite{
	aarts04,hernandez09,ramsteiner10,nguyen11,vanloenen19} 
Intriguingly, experimental measurement of the
stiffness coefficient of those same colloidal suspensions yield widely 
different results in different labs. Some authors find results 
in agreement  with the stiffness coefficient 
of the solid/liquid interface calculated in computer
simulations,\cite{nguyen11,vanloenen19} 
while others find results that differ as much as a factor of two.\cite{hernandez09,ramsteiner10}

Here we show that the surface tensions  of 'hard' colloid interfaces
obtained in experiments show a distinct gravitational dependence (Fig.1) that
is fully consistent with \Eq{gravity} and allows to reconcile experimental and 
theoretical results. The external
field dependence of the surface tension is explained bottom-up in terms of
an improved interface Hamiltonian which provides corrections to the fundamental equation of capillarity theory and
whose growth dynamics is closely
related to the Kardar-Parisi-Zhang model of deposition growth.

\begin{figure}[htb!]
   \centering
   \includegraphics[clip,scale=0.30]{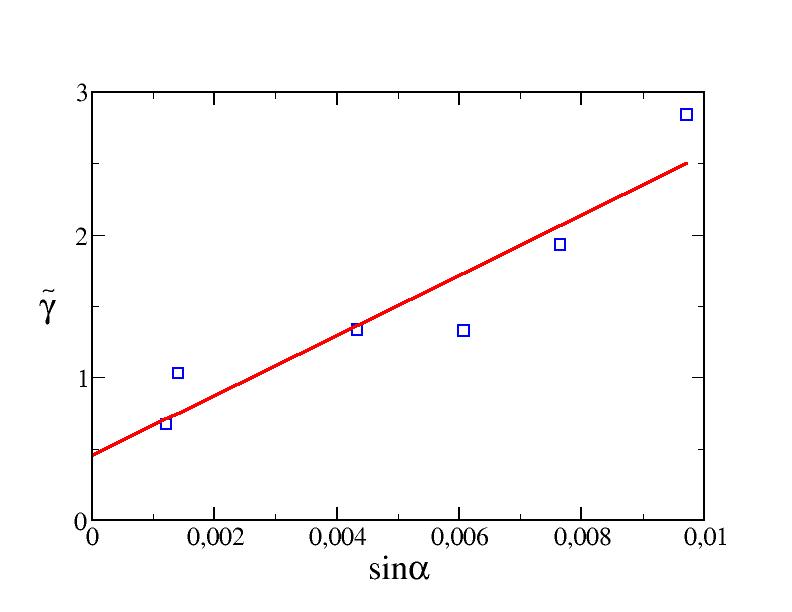}
     \caption{Stiffness coefficients of colloidal monolayers as
	  a function of gravity. The symbols are experimental
	  stiffness coefficients (in $10^{-16}$ J/m) from 
        Ref.\cite{thorneywork17} plotted
	  as a function of $\sin(\alpha)$, where $\alpha$ is
	  the tilt angle of the inclined monolayer. The straight
	  line is a least square fit under the assumption that
	  the stifness is a linear function of the gravity component
	  along the inclined plane, $G\sin(\alpha)$, as dictated in \Eq{gravity}. 
     }
\label{fig:gamma_v_sin}
\end{figure}

In order to illustrate the significance of \Eq{gravity}, 
we first consider 
experimental results by Thorneywork et al. for
two dimensional colloidal hard spheres.\cite{thorneywork17} 
These authors studied
the behavior of a tilted monolayer of colloids  deposited
on a glass surface.  By selecting the appropriate surface fraction of colloids
in the system, 
the monolayer phase separates into a liquid and an hexatic phase, 
with a well defined interface. The authors studied the interfacial
fluctuations by optical means,
and inferred directly the  stiffness coefficient
from the ratio of parallel to perpendicular fluctuations as predicted by
capillary wave theory in two dimensions. 

Surprisingly, independent realizations of the assembled monolayers yielded
significantly different stiffness coefficients. The authors attributed 
this to  different orientations of the solid hexatic phase
with respect to the interface position, and fitted
their results to a model of surface anisotropy with hexagonal symmetry.


Consider instead that the colloidal hard spheres are sufficiently massive
that the surface tension is afected by gravity. The tilt angle, $\alpha$, then
serves to tune the force of gravity along the inclined plane, 
and the component of the
field in the parallel direction to the monolayer plane is given
by $\Delta\rho\, G \sin(\alpha)$.
Plotting the surface stiffnesses reported in Ref.\cite{thorneywork17}
as a function of  $\sin(\alpha)$
clearly shows an increasing trend with
tilt angle, as predicted by \Eq{gravity} (Fig.\ref{fig:gamma_v_sin}). 
Performing a linear regression,
using $G=9.8~$ms$^{-2}$ and a surface density difference as reported in
Ref.\cite{thorneywork17}, provides a good fit, with a bulk correlation length of
$\xi=10~\mu$m, which is a reasonable value in view of the colloid's diameter,
$\sigma=2.79~\mu$m. Furthermore, the zero field stiffness, as obtained from the
linear fit to \Eq{gravity} yields $\tilde\gamma d/k_BT=0.031$, which is about one order of magnitude
smaller than the related liquid/solid stiffness coefficient in three dimensions,
in line with expectations. 

\begin{table}
   \begin{tabular}{c|ccc|cc}
	\hline
	\hline
   Orientation & $\beta\tilde\gamma\sigma^2$ & $\beta g''\sigma^4$ &
   $\beta(\tilde\gamma-\sigma^2 \Delta\rho {G})\sigma^2$ & $\beta\tilde\gamma\sigma^2$ &
   $\beta\gamma\sigma^2$ \\
	\hline
   (100) & 1.3 & 0.57 & 0.73 & 0.419 & 0.639 \\
   (100) & 1.1 & 0.49 & 0.61 & 0.419 & 0.639 \\
   (110)[$\bar{1}10$] & 1.0 & 0.37 & 0.63 & 0.769 & 0.616 \\
   (110)[001] & 1.0 & 0.37 & 0.63 & 0.401 & 0.616 \\
   (111) & 0.66 & 0.08 & 0.58 & 0.67 & - \\
	\hline
\end{tabular}
\caption{Stifness coefficients of hard sphere colloids with or without a
   gravitational field. The second and third columns provide results for
   hard sphere colloids under gravity from Ref.\cite{ramsteiner10}. The
   fourth column displays the gravity corrected result as described in Eq.1. The
   fifth {and sixth} column present computer simulation results 
   {for the stiffness coefficient $\tilde\gamma$ and the related surface tension $\gamma$} under zero gravity from
   Ref.\cite{hartel12}, except for the (111) plane, from Ref.\cite{davidchack06}. 
    Data 
   for the (111) plane correspond to a random stacking closed  packed crystal 
   both in experiments and simulations.
}
\label{3dcolloids}
\end{table}

A systematic study of surface properties with gravity  is not available
for 3-d hard sphere colloids. However, stiffness coefficients have
been measured for 3-d hard sphere colloids by Ramsteiner et
al.\cite{ramsteiner10} and Van Loenen et al.\cite{vanloenen19}.
Interestingly, Ramsteiner et al. performed experiments with a significant
gravity effect due to a mismatch of colloid and solvent density, and
found stiffness coefficients which
are about twice as large as those expected in computer simulations. On the
contrary, Van Loenen et al. chose a colloidal suspension with much closer
colloid-solvent density match, and found results  that are similar, albeit
somewhat smaller than theoretical expectations.  Indeed, the capillary wave 
analysis of Refs.\cite{ramsteiner10,vanloenen19}
allows to measure the effective gravitational damping,
$g''=\Delta\rho\, G$, directly from the spectrum of surface fluctuations. 
The results show that
$\Delta\rho G$ is of the same order of magnitude as 
$\tilde\gamma$ in the experiments by
Remsteiner et al., but is vanishingly small in those by Van Loenen et al.

According to \Eq{gravity}, the stiffness coefficients measured by 
Ramsteiner et al.
should therefore be significantly affected by gravity. We
can estimate the zero field stifness coefficients
of Ref.\cite{ramsteiner10}, as 
$\tilde\gamma_0 = \tilde\gamma(G) - \xi^2\Delta\rho\, G$, using
$\tilde\gamma(G)$ and $\Delta\rho\, G$ obtained
independently from their experiments, together with $\xi=\sigma$ as an order
of magnitude estimate for the interfacial width. The results
are displayed in Table~\ref{3dcolloids}, and compared with zero gravity results
obtained from computer simulations.\cite{hartel12} Despite some discrepancies, 
the table clearly shows that the gravity correction brings the experimental
results in much better agreement with computer simulations. 
Most strikingly, the stifness coefficient for the $(100)$ plane,
which has a large value of $g''$, differs by more that 260\% with 
zero gravity
results, and is brought to a 50\% discrepancy upon correction from \Eq{gravity}.
On the contrary, for the randomly stacked (111) plane, which has a small
value of $g''$, the experiments report stiffness coefficients that agree 
within 15\% with the zero gravity results.

The results shown here for the effect of gravity on interfacial
properties are in fact a special case of a more
general result regarding the dependence of stifness coefficients
on external fields, which reads:\cite{macdowell13,macdowell14,macdowell17}
\begin{equation}\label{eq:general}
 \gamma = \gamma_0 + \xi^2 g''
\end{equation} 
where $g''$ is the second derivative of the interface potential 
with respect to the interface position; while $\xi$ is
an empirical measure of the interfacial width, with similar order of
magnitude as the bulk correlation length.  
The accuracy of this result has been 
tested in computer simulation studies
for the special case of liquid films pinned on an inert 
substrate by van der Waals forces, 
where $g''$ decays as an inverse power law of the film 
width.\cite{macdowell13,benet14b,macdowell14,macdowell18}
For an interface pinned by gravity, on the contrary,
the interface potential is just equal to the gravitational potential energy,
$g=\frac{1}{2}\Delta\rho\, G h^2$, then  $g''=\Delta\rho\, G$ is a constant
and \Eq{general} becomes equal to \Eq{gravity}.

The result of \Eq{general} can be derived from an interface 
displacement model, assuming that the density of a corrugated 
interface, $\rho({\bf r})$ is a function of the perpendicular 
distance away from the
interface location:\cite{macdowell17}
\begin{equation}\label{eq:density}
   \rho({\bf r })=\rho_{\pi}\left
(\frac{z-\hh({\x)}}{\sqrt{1+(\nabla\hh)^2}} \right )
\end{equation}
where $\rho({\bf r})$ is the fluid's density for a given
realization of the fluctuations, $\rho_{\pi}(z)$ is the mean field density
of a flat interface, $\hh({\bf x})$ is the interface position
in the Monge representation, ${\bf x}$ is a point on a reference plane
oriented parallel to the average interface position, and $z$ is the
perpendicular distance to that plane. This expression shows that the
density profile of a corrugated interface depends not only on $\hh({\bf x})$,
but also on $\nabla\hh({\bf x})$, which is a simple way to convey the non-locality
of corrugated interfaces on the interface position $\hh({\bf x})$.\cite{parry04}

This  assumption, which has been explored in a number
of studies,\cite{davis77,mecke99b} has 
been shown to be far more accurate than the standard interface
displacement model $\rho({\bf r })=\rho_{\pi}(z-\hh({\bf x}))$ for the description of
sessile droplets barely a few molecular diameters away from the substrate.\cite{nold16}
In fact, using the familiar microscopic van der Waals theory of interfaces,\cite{rowlinson82b} 
\Eq{density} yields exactly the coarse-grained interface Hamiltonian model:\cite{davis77,macdowell17},
 \begin{equation}
 H[\hh] =  \gamma_0 \int \sqrt{ 1 + \left( \nabla\hh \right)^2 } d{\bf x}
 \end{equation} 
 

{In the presence of an external field, the free energy functional can
become far more complex, as the intrinsic density profile $\rho_{\pi}(z)$ in
\Eq{density} is modified by the field.\cite{bernardino09}. However,
already to zero order in the density profile, there appear interesting
corrections, whose significance has not been widely recognized. }
Indeed,  assuming a local potential
$V(z)$ acts on the system, one finds:\cite{benet14b}
\begin{widetext}
\begin{equation}\label{eq:functional}
    H[\hh] = \int d{\bf x} \left [ \int d z\, V(z) \rho_{\pi}\left(\frac{z-\hh(\x)}{\sqrt{1
		    +
	 (\nabla \hh)^2}}\right) +    \gamma_0 \, \sqrt{1+(\nabla
	 \hh)^2}
	   - \Delta p\,\hh(\x) \right ] 
\end{equation} 
\end{widetext}
{where $\Delta p$ stands for the Laplace pressure difference across the interface} and we have purposely avoided explicit integration of the external
field over the volume,  which cannot be readily performed without additional
approximations.\cite{pahlavan18,benet14b}  In the classical theory, this integral is equated to the interface potential of a flat interface evaluated at the local
interface position,  $g(h)$. Instead, by seeking for the extremal of
the free energy prior to  integration of $V(z)$ over volume,
we find a new equilibrium condition for liquid films which goes beyond
the traditional capillary approximation:
{
\begin{equation}\label{eq:extremal}
  \frac{\tilde\Pi(\hh,\hh_{\x})}{\sqrt{1 + 	\hh_{\x}^2}} + \Delta p = -\frac{d}{d\x}\left ( \frac{\gamma_0 \hh_{\x}}{\sqrt{1 + 	\hh_{\x}^2}} + \frac{\Delta \tilde\gamma(\hh,\hh_{\x})  \hh{_{\x}}}{(1 + 	\hh_{\x}^2)^{3/2}}
\right )
\end{equation}
where $\tilde\Pi(\hh,\hh_{\x})$ is the disjoining pressure, $\tilde\Delta\gamma(\hh,\hh_{\x})$ is the extrinsic surface tension due to the external field and $\hh_{\x}$ is used here as shorthand for $\nabla \hh$. The tilde on  $\Pi$ and $\Delta\gamma$ denotes that these objects are actually complicated non-local  functionals of the film profile, as conveyed by their explicit dependence on  the film gradient. 
}

{
In practice, for the
usual case where the external field $V(z)$ varies smoothly on the scale of the interfacial width, the $\hh$ and $\hh_{\x}$ dependencies 
in \rechange{$\tilde \Pi$} conveniently factor out as
$\tilde\Pi(\hh,\hh_{\x})\approx\sqrt{1 + 	\hh_{\x}^2}\,\Pi(\hh)$ with $\Pi(\hh)$
the disjoining pressure of a planar interface (this simplification
was overlooked in Ref.\cite{benet14b}, and lead to a linearized extremal condition that is in error).
\rechange{Using this result and assuming the limit of small gradients, such that $\Delta\tilde\gamma\to\Delta\gamma(\hh)$, 
with  $\Delta\gamma(\hh)=\xi^2 g''(h)$,
\Eq{extremal} now becomes a non-linear differential equation (Suppemental Material):}
\rechange{
\begin{equation}\label{eq:idrjg}
\Pi(\hh) + \Delta p = -\frac{d}{d\x}\left (\frac{}{}{\gamma(\hh) \hh_{\x}} \right )
\end{equation}
}
}

{
Neglecting the $\hh$ dependence of $\gamma(\hh)$, \Eq{idrjg} recovers   the traditional Derjaguin or augmented Young-Laplace equation, which is widely used to predict the equilibrium shape and spreading dynamics of sessile 
droplets and capillary bridges.\cite{degennes85,davidovitch05,churaev88,degennes04,starov09,yin17,pahlavan18,duran-olivencia19,zhang20,saiseau22} However, \Eq{general} shows that  corrections to the  surface tension may become important in the neighborhood of the three phase contact region, where $g''(\hh)$ becomes large. 
}

{
To see this, consider 
the first integral of \Eq{idrjg}, which, to leading order in $g(\hh)/\gamma_0$ is given as (Supplemental Material):
\begin{equation}\label{eq:1stint}
\hh_{\x}^2 = \frac{2(g(\hh)-g(\hh_e)) + \frac{1}{2}\frac{\xi^2}{\gamma_0}\Pi^2(\hh)}{\gamma_0 + \xi^2 g''(\hh)}
\end{equation}
where $\hh_e$ is the equilibrium film thickness of a flat film.}

{
Away from the three phase contact line, $\Pi^2(\hh)$ and $g''(\hh)$ decay to zero faster than $g(\hh)$ does, and the above result recovers exactly the first integral of the Derjaguin equation.\cite{churaev88,degennes04,starov09}  In the neighborhood of the substrate, however, \Eq{1stint} provides significant corrections and dictates  deviations of the film profile $\hh_{x}\approx \theta$  from the macroscopic contact angle, $\theta\approx\sqrt{-2g(\hh_e)/\gamma_0}$. In practice, since  $\Pi^2(\hh)$ usually decays faster than $g''(\hh)$, the qualitative change may be assessed by ignoring $\Pi^2(\hh)$ altogether.
}

{
As an explicit example, consider a model interface potential exhibiting incomplete wetting, with an  equilibrium film thickness of about two correlation lengths, and a contact angle of about $\theta=40$ degrees (Supplemental Material). Solving \Eq{1stint} for this model  under the appropriate boundary conditions, provides the film profile of
a cylindrical liquid droplet (Figure 2). Away from the substrate, $g(\hh)$ is dominated by the long rage dispersion tail, and  $\Delta\gamma$ provides a small positive correction to $\gamma_0$ which has a negligible effect in the film profile. However, as the profile approaches the substrate, $\Delta\gamma$ becomes large and negative (Figure 3-Inset). As a result, the slope of $\hh(\x)$ becomes larger than predicted by the Derjaguin equation, and the  film profile falls sharply towards the substrate. Eventually, as $\hh$ approaches the equilibrium film thickness,  $\Delta\gamma$ becomes positive again and the asymptotic approach towards $\hh_e$ becomes smoother than that predicted by the Derjaguin equation (Figure 2). Therefore, the corrections due to the $\hh$ dependence of the surface tension can become noticeable within a range of a few correlation lengths.
}

\begin{figure}[htb!]
	\centering
	\includegraphics[clip,scale=0.30]{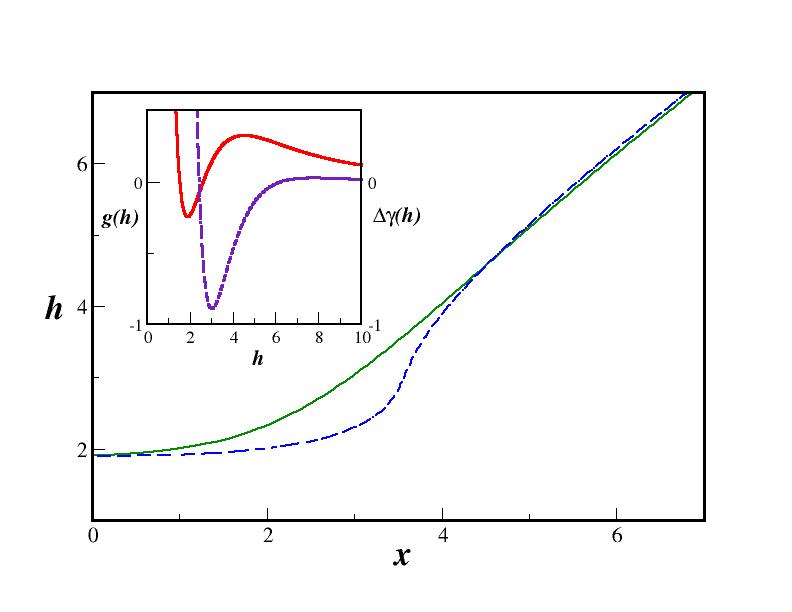}
	\caption{{Shape of liquid droplet approaching the three phase contact
line. The green (full) line is the predicted drop profile according to the Derjaguin
equation, while the blue (dashed) line corresponds to predictions from
\Eq{1stint}. The inset shows the model interface potential employed (red full
line, left axis) and $\Delta\gamma(\hh)$ (violet  dashed line, right axis). 
The lengthscale of both figures is given in units of the correlation length, 
and the surface energy scale in units of $\gamma_0$.
}}
	\label{fig:droplet}
\end{figure}


The  improved functional, \Eq{functional}, also has interesting implications for the dynamics of interfaces. Indeed, we notice that in the small slope
approximation, the non-conserved
gradient driven dynamics of the functional in \Eq{functional} yields
readily a deterministic non-linear differential equation for the deposition dynamics of a gas at coexistence ($\Delta p=0$):
{
\begin{equation}\label{eq:kpzm}
    \frac{\partial \hh}{\partial t} =  
    \Pi(h)  + \gamma(\hh) \frac{d^2\hh}{dx^2} +  
	    \gamma'(h)  \left( \frac{d\hh}{dx} \right )^2 
\end{equation}
} 
Adding a random white noise term, this result becomes a non-linear stochastic
growth model which  may be viewed as a generalization of the celebrated
Kardar-Parisi-Zhang equation (KPZ) of deposition growth.\cite{kardar86} Here it is shown transparently that the non-linear term may be obtained from an equilibrium
free energy functional, an issue that has been a matter of some debate (c.f.
Ref.\cite{wio09,wio22} for a review). The bottom-up derivation makes explicit the origin of the phenomenological coefficients, and shows that they are not fully independent.  

%
%

For a  thin adsorbed film above the roughening transition,
the interface potential decreases with distance, and \Eq{kpzm} yields a KPZ equation with
a monotonously decaying driving and variable coefficients of the linear and quadratic terms. When the adsorbed film becomes thick enough (i.e. such as in an ordinary fluid
interface), the effect of the adsorbent's
external field is negligible, $g(h)\to 0$, and both the driving term and the quadratic coefficient  
vanish altogether, leading to a standard result of 
deposition growth on a fluid interface.\cite{thiele10} Therefore, \Eq{kpzm} predicts
for growth of rough films a smooth crossover from a solid-like to a liquid like
deposition mechanism as the film grows. 
On the contrary,   
for a film growing below its roughening transition
(as is the case of epitaxial  growth), $g(h)$ is oscillatory,\cite{chernov88}.
In this case, \Eq{kpzm} recovers the sine gordon-model of crystal growth,\cite{saito80,moro01b} albeit with a quadratic correction which
resembles  the KPZ equation.
Most interestingly, the coefficients are oscillatory, and the quadratic coefficient periodically changes sign.
These features anticipate a rich  behavior not
predicted by the strict KPZ equation alone, and challenges the view that
the constant coefficient model universally describes the long scale behavior
of growing interfaces.
.%

In summary, we provide compelling evidence of the influence of gravity
on measured surface tensions. The interfacial
Hamiltonian required to explain this behavior 
provides corrections to the fundamental laws of 
capillarity theory and thin film
deposition under external fields, with {potential} implications 
in a wide range of applications. 



\begin{acknowledgments}
{	I am indebted to an anonimous reviewer for invaluable comments}. I would also like to thank J\"urgen Horbach, Ram\'on Gonz\'alez Rubio and Horacio Wio for helpful discussions.  Funding  from the Spanish Agencia Estatal de Investigaci\'on under research  grant PID2020-115722GB-C21 is greatefully acknowledged.
\end{acknowledgments}


\begin{thebibliography}{41}%
\makeatletter
\providecommand \@ifxundefined [1]{%
 \@ifx{#1\undefined}
}%
\providecommand \@ifnum [1]{%
 \ifnum #1\expandafter \@firstoftwo
 \else \expandafter \@secondoftwo
 \fi
}%
\providecommand \@ifx [1]{%
 \ifx #1\expandafter \@firstoftwo
 \else \expandafter \@secondoftwo
 \fi
}%
\providecommand \natexlab [1]{#1}%
\providecommand \enquote  [1]{``#1''}%
\providecommand \bibnamefont  [1]{#1}%
\providecommand \bibfnamefont [1]{#1}%
\providecommand \citenamefont [1]{#1}%
\providecommand \href@noop [0]{\@secondoftwo}%
\providecommand \href [0]{\begingroup \@sanitize@url \@href}%
\providecommand \@href[1]{\@@startlink{#1}\@@href}%
\providecommand \@@href[1]{\endgroup#1\@@endlink}%
\providecommand \@sanitize@url [0]{\catcode `\\12\catcode `\$12\catcode
  `\&12\catcode `\#12\catcode `\^12\catcode `\_12\catcode `\%12\relax}%
\providecommand \@@startlink[1]{}%
\providecommand \@@endlink[0]{}%
\providecommand \url  [0]{\begingroup\@sanitize@url \@url }%
\providecommand \@url [1]{\endgroup\@href {#1}{\urlprefix }}%
\providecommand \urlprefix  [0]{URL }%
\providecommand \Eprint [0]{\href }%
\providecommand \doibase [0]{https://doi.org/}%
\providecommand \selectlanguage [0]{\@gobble}%
\providecommand \bibinfo  [0]{\@secondoftwo}%
\providecommand \bibfield  [0]{\@secondoftwo}%
\providecommand \translation [1]{[#1]}%
\providecommand \BibitemOpen [0]{}%
\providecommand \bibitemStop [0]{}%
\providecommand \bibitemNoStop [0]{.\EOS\space}%
\providecommand \EOS [0]{\spacefactor3000\relax}%
\providecommand \BibitemShut  [1]{\csname bibitem#1\endcsname}%
\let\auto@bib@innerbib\@empty
\bibitem [{\citenamefont {Buff}\ \emph {et~al.}(1965)\citenamefont {Buff},
  \citenamefont {Lovett},\ and\ \citenamefont {Stillinger}}]{buff65}%
  \BibitemOpen
  \bibfield  {author} {\bibinfo {author} {\bibfnamefont {F.~P.}\ \bibnamefont
  {Buff}}, \bibinfo {author} {\bibfnamefont {R.~A.}\ \bibnamefont {Lovett}},\
  and\ \bibinfo {author} {\bibfnamefont {F.~H.}\ \bibnamefont {Stillinger}},\
  }\bibfield  {title} {\bibinfo {title} {Interfacial density profile for fluids
  in the critical region},\ }\href {https://doi.org/10.1103/PhysRevLett.15.621}
  {\bibfield  {journal} {\bibinfo  {journal} {Phys. Rev. Lett.}\ }\textbf
  {\bibinfo {volume} {15}},\ \bibinfo {pages} {621} (\bibinfo {year}
  {1965})}\BibitemShut {NoStop}%
\bibitem [{\citenamefont {Zittartz}(1967)}]{zittartz67}%
  \BibitemOpen
  \bibfield  {author} {\bibinfo {author} {\bibfnamefont {J.}~\bibnamefont
  {Zittartz}},\ }\bibfield  {title} {\bibinfo {title} {Microscopic approach to
  interfacial structure in ising-like ferromagnets},\ }\href
  {https://doi.org/10.1103/PhysRev.154.529} {\bibfield  {journal} {\bibinfo
  {journal} {Phys. Rev.}\ }\textbf {\bibinfo {volume} {154}},\ \bibinfo {pages}
  {529} (\bibinfo {year} {1967})}\BibitemShut {NoStop}%
\bibitem [{\citenamefont {Jasnow}(1984)}]{jasnow84}%
  \BibitemOpen
  \bibfield  {author} {\bibinfo {author} {\bibfnamefont {D.}~\bibnamefont
  {Jasnow}},\ }\bibfield  {title} {\bibinfo {title} {Critical phenomena at
  interfaces},\ }\href@noop {} {\bibfield  {journal} {\bibinfo  {journal} {Rep.
  Prog. Phys.}\ }\textbf {\bibinfo {volume} {47}},\ \bibinfo {pages} {1059}
  (\bibinfo {year} {1984})}\BibitemShut {NoStop}%
\bibitem [{\citenamefont {Rowlinson}\ and\ \citenamefont
  {Widom}(1982)}]{rowlinson82b}%
  \BibitemOpen
  \bibfield  {author} {\bibinfo {author} {\bibfnamefont {J.}~\bibnamefont
  {Rowlinson}}\ and\ \bibinfo {author} {\bibfnamefont {B.}~\bibnamefont
  {Widom}},\ }\href@noop {} {\emph {\bibinfo {title} {Molecular Theory of
  Capillarity}}}\ (\bibinfo  {publisher} {Clarendon},\ \bibinfo {address}
  {Oxford},\ \bibinfo {year} {1982})\BibitemShut {NoStop}%
\bibitem [{\citenamefont {MacDowell}\ \emph {et~al.}(2013)\citenamefont
  {MacDowell}, \citenamefont {Benet},\ and\ \citenamefont
  {Katcho}}]{macdowell13}%
  \BibitemOpen
  \bibfield  {author} {\bibinfo {author} {\bibfnamefont {L.~G.}\ \bibnamefont
  {MacDowell}}, \bibinfo {author} {\bibfnamefont {J.}~\bibnamefont {Benet}},\
  and\ \bibinfo {author} {\bibfnamefont {N.~A.}\ \bibnamefont {Katcho}},\
  }\bibfield  {title} {\bibinfo {title} {Capillary fluctuations and
  film-height-dependent surface tension of an adsorbed liquid film},\ }\href
  {https://doi.org/10.1103/PhysRevLett.111.047802} {\bibfield  {journal}
  {\bibinfo  {journal} {Phys. Rev. Lett.}\ }\textbf {\bibinfo {volume} {111}},\
  \bibinfo {pages} {047802} (\bibinfo {year} {2013})}\BibitemShut {NoStop}%
\bibitem [{\citenamefont {Pusey}\ and\ \citenamefont {van
  Mengen}(1986)}]{pusey86}%
  \BibitemOpen
  \bibfield  {author} {\bibinfo {author} {\bibfnamefont {P.}~\bibnamefont
  {Pusey}}\ and\ \bibinfo {author} {\bibfnamefont {W.}~\bibnamefont {van
  Mengen}},\ }\bibfield  {title} {\bibinfo {title} {Phase behavior of
  concentrated solutions of nearly hard colloidal spheres},\ }\href@noop {}
  {\bibfield  {journal} {\bibinfo  {journal} {Nature}\ }\textbf {\bibinfo
  {volume} {320}},\ \bibinfo {pages} {340} (\bibinfo {year}
  {1986})}\BibitemShut {NoStop}%
\bibitem [{\citenamefont {van Blaaderen}\ and\ \citenamefont
  {Wiltzius}(1995)}]{vanblaaderen95}%
  \BibitemOpen
  \bibfield  {author} {\bibinfo {author} {\bibfnamefont {A.}~\bibnamefont {van
  Blaaderen}}\ and\ \bibinfo {author} {\bibfnamefont {P.}~\bibnamefont
  {Wiltzius}},\ }\bibfield  {title} {\bibinfo {title} {Real-space structure of
  colloidal hard-sphere glasses},\ }\href
  {https://doi.org/10.1126/science.270.5239.1177} {\bibfield  {journal}
  {\bibinfo  {journal} {Science}\ }\textbf {\bibinfo {volume} {270}},\ \bibinfo
  {pages} {1177} (\bibinfo {year} {1995})},\ \Eprint
  {https://arxiv.org/abs/https://www.science.org/doi/pdf/10.1126/science.270.5239.1177}
  {https://www.science.org/doi/pdf/10.1126/science.270.5239.1177} \BibitemShut
  {NoStop}%
\bibitem [{\citenamefont {Aarts}\ \emph {et~al.}(2004)\citenamefont {Aarts},
  \citenamefont {Schmidtt},\ and\ \citenamefont {Lekkerkerker}}]{aarts04}%
  \BibitemOpen
  \bibfield  {author} {\bibinfo {author} {\bibfnamefont {D.~G.}\ \bibnamefont
  {Aarts}}, \bibinfo {author} {\bibfnamefont {M.}~\bibnamefont {Schmidtt}},\
  and\ \bibinfo {author} {\bibfnamefont {H.~N.~K.}\ \bibnamefont
  {Lekkerkerker}},\ }\bibfield  {title} {\bibinfo {title} {Direct observation
  of thermal capillary waves},\ }\href@noop {} {\bibfield  {journal} {\bibinfo
  {journal} {Science}\ }\textbf {\bibinfo {volume} {304}},\ \bibinfo {pages}
  {847} (\bibinfo {year} {2004})}\BibitemShut {NoStop}%
\bibitem [{\citenamefont {Hern\'andez-Guzm\'an}\ and\ \citenamefont
  {Weeks}(2009)}]{hernandez09}%
  \BibitemOpen
  \bibfield  {author} {\bibinfo {author} {\bibfnamefont {J.}~\bibnamefont
  {Hern\'andez-Guzm\'an}}\ and\ \bibinfo {author} {\bibfnamefont {E.~R.}\
  \bibnamefont {Weeks}},\ }\bibfield  {title} {\bibinfo {title} {The
  equilibrium intrinsic crystal-liquid interface of colloids},\ }\href@noop {}
  {\bibfield  {journal} {\bibinfo  {journal} {Proc. Natl. Acad. Sci. U.S.A.}\
  }\textbf {\bibinfo {volume} {106}},\ \bibinfo {pages} {15198} (\bibinfo
  {year} {2009})}\BibitemShut {NoStop}%
\bibitem [{\citenamefont {Ramsteiner}\ \emph {et~al.}(2010)\citenamefont
  {Ramsteiner}, \citenamefont {Weitz},\ and\ \citenamefont
  {Spaepen}}]{ramsteiner10}%
  \BibitemOpen
  \bibfield  {author} {\bibinfo {author} {\bibfnamefont {I.~B.}\ \bibnamefont
  {Ramsteiner}}, \bibinfo {author} {\bibfnamefont {D.~A.}\ \bibnamefont
  {Weitz}},\ and\ \bibinfo {author} {\bibfnamefont {F.}~\bibnamefont
  {Spaepen}},\ }\bibfield  {title} {\bibinfo {title} {Stiffness of the
  crystal-liquid interface in a hard-sphere colloidal system measured from
  capillary fluctuations},\ }\href {https://doi.org/10.1103/PhysRevE.82.041603}
  {\bibfield  {journal} {\bibinfo  {journal} {Phys. Rev. E}\ }\textbf {\bibinfo
  {volume} {82}},\ \bibinfo {pages} {041603} (\bibinfo {year}
  {2010})}\BibitemShut {NoStop}%
\bibitem [{\citenamefont {Nguyen}\ \emph {et~al.}(2011)\citenamefont {Nguyen},
  \citenamefont {Hu},\ and\ \citenamefont {Schall}}]{nguyen11}%
  \BibitemOpen
  \bibfield  {author} {\bibinfo {author} {\bibfnamefont {V.~D.}\ \bibnamefont
  {Nguyen}}, \bibinfo {author} {\bibfnamefont {Z.}~\bibnamefont {Hu}},\ and\
  \bibinfo {author} {\bibfnamefont {P.}~\bibnamefont {Schall}},\ }\bibfield
  {title} {\bibinfo {title} {Single crystal growth and anisotropic
  crystal-fluid interfacial free energy in soft colloidal systems},\ }\href
  {https://doi.org/10.1103/PhysRevE.84.011607} {\bibfield  {journal} {\bibinfo
  {journal} {Phys. Rev. E}\ }\textbf {\bibinfo {volume} {84}},\ \bibinfo
  {pages} {011607} (\bibinfo {year} {2011})}\BibitemShut {NoStop}%
\bibitem [{\citenamefont {van Loenen}\ \emph {et~al.}(2019)\citenamefont {van
  Loenen}, \citenamefont {Kodger}, \citenamefont {Padston}, \citenamefont
  {Nawar}, \citenamefont {Schall},\ and\ \citenamefont
  {Spaepen}}]{vanloenen19}%
  \BibitemOpen
  \bibfield  {author} {\bibinfo {author} {\bibfnamefont {S.~Z.}\ \bibnamefont
  {van Loenen}}, \bibinfo {author} {\bibfnamefont {T.~E.}\ \bibnamefont
  {Kodger}}, \bibinfo {author} {\bibfnamefont {E.~A.}\ \bibnamefont {Padston}},
  \bibinfo {author} {\bibfnamefont {S.}~\bibnamefont {Nawar}}, \bibinfo
  {author} {\bibfnamefont {P.}~\bibnamefont {Schall}},\ and\ \bibinfo {author}
  {\bibfnamefont {F.}~\bibnamefont {Spaepen}},\ }\bibfield  {title} {\bibinfo
  {title} {Measurement of the stiffness of hard-sphere colloidal crystal-liquid
  interfaces},\ }\href {https://doi.org/10.1103/PhysRevMaterials.3.085605}
  {\bibfield  {journal} {\bibinfo  {journal} {Phys. Rev. Mater.}\ }\textbf
  {\bibinfo {volume} {3}},\ \bibinfo {pages} {085605} (\bibinfo {year}
  {2019})}\BibitemShut {NoStop}%
\bibitem [{\citenamefont {Thorneywork}\ \emph {et~al.}(2017)\citenamefont
  {Thorneywork}, \citenamefont {Abbott}, \citenamefont {Aarts},\ and\
  \citenamefont {Dullens}}]{thorneywork17}%
  \BibitemOpen
  \bibfield  {author} {\bibinfo {author} {\bibfnamefont {A.~L.}\ \bibnamefont
  {Thorneywork}}, \bibinfo {author} {\bibfnamefont {J.~L.}\ \bibnamefont
  {Abbott}}, \bibinfo {author} {\bibfnamefont {D.~G. A.~L.}\ \bibnamefont
  {Aarts}},\ and\ \bibinfo {author} {\bibfnamefont {R.~P.~A.}\ \bibnamefont
  {Dullens}},\ }\bibfield  {title} {\bibinfo {title} {Two-dimensional melting
  of colloidal hard spheres},\ }\href
  {https://doi.org/10.1103/PhysRevLett.118.158001} {\bibfield  {journal}
  {\bibinfo  {journal} {Phys. Rev. Lett.}\ }\textbf {\bibinfo {volume} {118}},\
  \bibinfo {pages} {158001} (\bibinfo {year} {2017})}\BibitemShut {NoStop}%
\bibitem [{\citenamefont {H\"artel}\ \emph {et~al.}(2012)\citenamefont
  {H\"artel}, \citenamefont {Oettel}, \citenamefont {Rozas}, \citenamefont
  {Egelhaaf}, \citenamefont {Horbach},\ and\ \citenamefont
  {L\"owen}}]{hartel12}%
  \BibitemOpen
  \bibfield  {author} {\bibinfo {author} {\bibfnamefont {A.}~\bibnamefont
  {H\"artel}}, \bibinfo {author} {\bibfnamefont {M.}~\bibnamefont {Oettel}},
  \bibinfo {author} {\bibfnamefont {R.~E.}\ \bibnamefont {Rozas}}, \bibinfo
  {author} {\bibfnamefont {S.~U.}\ \bibnamefont {Egelhaaf}}, \bibinfo {author}
  {\bibfnamefont {J.}~\bibnamefont {Horbach}},\ and\ \bibinfo {author}
  {\bibfnamefont {H.}~\bibnamefont {L\"owen}},\ }\bibfield  {title} {\bibinfo
  {title} {Tension and stiffness of the hard sphere crystal-fluid interface},\
  }\href {https://doi.org/10.1103/PhysRevLett.108.226101} {\bibfield  {journal}
  {\bibinfo  {journal} {Phys. Rev. Lett.}\ }\textbf {\bibinfo {volume} {108}},\
  \bibinfo {pages} {226101} (\bibinfo {year} {2012})}\BibitemShut {NoStop}%
\bibitem [{\citenamefont {Davidchack}\ \emph {et~al.}(2006)\citenamefont
  {Davidchack}, \citenamefont {Morris},\ and\ \citenamefont
  {Laird}}]{davidchack06}%
  \BibitemOpen
  \bibfield  {author} {\bibinfo {author} {\bibfnamefont {R.~L.}\ \bibnamefont
  {Davidchack}}, \bibinfo {author} {\bibfnamefont {J.~R.}\ \bibnamefont
  {Morris}},\ and\ \bibinfo {author} {\bibfnamefont {B.~B.}\ \bibnamefont
  {Laird}},\ }\bibfield  {title} {\bibinfo {title} {The anisotropic hard-sphere
  crystal-melt interfacial free energy from fluctuations},\ }\href
  {https://doi.org/10.1063/1.2338303} {\bibfield  {journal} {\bibinfo
  {journal} {J. Chem. Phys.}\ }\textbf {\bibinfo {volume} {125}},\ \bibinfo
  {eid} {094710} (\bibinfo {year} {2006})}\BibitemShut {NoStop}%
\bibitem [{\citenamefont {MacDowell}\ \emph {et~al.}(2014)\citenamefont
  {MacDowell}, \citenamefont {Benet}, \citenamefont {Katcho},\ and\
  \citenamefont {Palanco}}]{macdowell14}%
  \BibitemOpen
  \bibfield  {author} {\bibinfo {author} {\bibfnamefont {L.~G.}\ \bibnamefont
  {MacDowell}}, \bibinfo {author} {\bibfnamefont {J.}~\bibnamefont {Benet}},
  \bibinfo {author} {\bibfnamefont {N.~A.}\ \bibnamefont {Katcho}},\ and\
  \bibinfo {author} {\bibfnamefont {J.~M.}\ \bibnamefont {Palanco}},\
  }\bibfield  {title} {\bibinfo {title} {Disjoining pressure and the
  film-height-dependent surface tension of thin liquid films: New insight from
  capillary wave fluctuations},\ }\href
  {https://doi.org/http://dx.doi.org/10.1016/j.cis.2013.11.003} {\bibfield
  {journal} {\bibinfo  {journal} {Adv. Colloid Interface Sci.}\ }\textbf
  {\bibinfo {volume} {206}},\ \bibinfo {pages} {150} (\bibinfo {year}
  {2014})}\BibitemShut {NoStop}%
\bibitem [{\citenamefont {MacDowell}(2017)}]{macdowell17}%
  \BibitemOpen
  \bibfield  {author} {\bibinfo {author} {\bibfnamefont {L.~G.}\ \bibnamefont
  {MacDowell}},\ }\bibfield  {title} {\bibinfo {title} {Capillary wave theory
  of adsorbed liquid films and the structure of the liquid-vapor interface},\
  }\href@noop {} {\bibfield  {journal} {\bibinfo  {journal} {Phys. Rev. E}\
  }\textbf {\bibinfo {volume} {96}},\ \bibinfo {pages} {022801} (\bibinfo
  {year} {2017})}\BibitemShut {NoStop}%
\bibitem [{\citenamefont {Benet}\ \emph {et~al.}(2014)\citenamefont {Benet},
  \citenamefont {Palanco}, \citenamefont {Sanz},\ and\ \citenamefont
  {MacDowell}}]{benet14b}%
  \BibitemOpen
  \bibfield  {author} {\bibinfo {author} {\bibfnamefont {J.}~\bibnamefont
  {Benet}}, \bibinfo {author} {\bibfnamefont {J.~G.}\ \bibnamefont {Palanco}},
  \bibinfo {author} {\bibfnamefont {E.}~\bibnamefont {Sanz}},\ and\ \bibinfo
  {author} {\bibfnamefont {L.~G.}\ \bibnamefont {MacDowell}},\ }\bibfield
  {title} {\bibinfo {title} {Disjoining pressure, healing distance, and film
  height dependent surface tension of thin wetting films},\ }\href@noop {}
  {\bibfield  {journal} {\bibinfo  {journal} {J. Phys. Chem. C}\ }\textbf
  {\bibinfo {volume} {118}},\ \bibinfo {pages} {22079} (\bibinfo {year}
  {2014})}\BibitemShut {NoStop}%
\bibitem [{\citenamefont {MacDowell}\ \emph {et~al.}(2018)\citenamefont
  {MacDowell}, \citenamefont {Llombart}, \citenamefont {Benet}, \citenamefont
  {Palanco},\ and\ \citenamefont {Guerrero-Martinez}}]{macdowell18}%
  \BibitemOpen
  \bibfield  {author} {\bibinfo {author} {\bibfnamefont {L.~G.}\ \bibnamefont
  {MacDowell}}, \bibinfo {author} {\bibfnamefont {P.}~\bibnamefont {Llombart}},
  \bibinfo {author} {\bibfnamefont {J.}~\bibnamefont {Benet}}, \bibinfo
  {author} {\bibfnamefont {J.~G.}\ \bibnamefont {Palanco}},\ and\ \bibinfo
  {author} {\bibfnamefont {A.}~\bibnamefont {Guerrero-Martinez}},\ }\bibfield
  {title} {\bibinfo {title} {Nanocapillarity and liquid bridge-mediated force
  between colloidal nanoparticles},\ }\href
  {https://doi.org/10.1021/acsomega.7b01650} {\bibfield  {journal} {\bibinfo
  {journal} {ACS Omega}\ }\textbf {\bibinfo {volume} {3}},\ \bibinfo {pages}
  {112} (\bibinfo {year} {2018})},\ \Eprint
  {https://arxiv.org/abs/http://dx.doi.org/10.1021/acsomega.7b01650}
  {http://dx.doi.org/10.1021/acsomega.7b01650} \BibitemShut {NoStop}%
\bibitem [{\citenamefont {Parry}\ \emph {et~al.}(2004)\citenamefont {Parry},
  \citenamefont {Romero-Enrique},\ and\ \citenamefont {Lazarides}}]{parry04}%
  \BibitemOpen
  \bibfield  {author} {\bibinfo {author} {\bibfnamefont {A.~O.}\ \bibnamefont
  {Parry}}, \bibinfo {author} {\bibfnamefont {J.~M.}\ \bibnamefont
  {Romero-Enrique}},\ and\ \bibinfo {author} {\bibfnamefont {A.}~\bibnamefont
  {Lazarides}},\ }\bibfield  {title} {\bibinfo {title} {Nonlocality and
  short-range wetting phenomena},\ }\href
  {https://doi.org/10.1103/PhysRevLett.93.086104} {\bibfield  {journal}
  {\bibinfo  {journal} {Phys. Rev. Lett.}\ }\textbf {\bibinfo {volume} {93}},\
  \bibinfo {pages} {086104} (\bibinfo {year} {2004})}\BibitemShut {NoStop}%
\bibitem [{\citenamefont {Davis}(1977)}]{davis77}%
  \BibitemOpen
  \bibfield  {author} {\bibinfo {author} {\bibfnamefont {H.~T.}\ \bibnamefont
  {Davis}},\ }\bibfield  {title} {\bibinfo {title} {Capillary waves and the
  mean field theory of interfaces},\ }\href
  {https://doi.org/http://dx.doi.org/10.1063/1.435301} {\bibfield  {journal}
  {\bibinfo  {journal} {J. Chem. Phys.}\ }\textbf {\bibinfo {volume} {67}},\
  \bibinfo {pages} {3636} (\bibinfo {year} {1977})}\BibitemShut {NoStop}%
\bibitem [{\citenamefont {Mecke}\ and\ \citenamefont
  {Dietrich}(1999)}]{mecke99b}%
  \BibitemOpen
  \bibfield  {author} {\bibinfo {author} {\bibfnamefont {K.~R.}\ \bibnamefont
  {Mecke}}\ and\ \bibinfo {author} {\bibfnamefont {S.}~\bibnamefont
  {Dietrich}},\ }\bibfield  {title} {\bibinfo {title} {Effective hamiltonian
  for liquid-vapor interfaces},\ }\href
  {https://doi.org/10.1103/PhysRevE.59.6766} {\bibfield  {journal} {\bibinfo
  {journal} {Phys. Rev. E}\ }\textbf {\bibinfo {volume} {59}},\ \bibinfo
  {pages} {6766} (\bibinfo {year} {1999})}\BibitemShut {NoStop}%
\bibitem [{\citenamefont {Nold}\ \emph {et~al.}(2018)\citenamefont {Nold},
  \citenamefont {MacDowell}, \citenamefont {Sibley}, \citenamefont {Goddard},\
  and\ \citenamefont {Kalliadasis}}]{nold16}%
  \BibitemOpen
  \bibfield  {author} {\bibinfo {author} {\bibfnamefont {A.}~\bibnamefont
  {Nold}}, \bibinfo {author} {\bibfnamefont {L.~G.}\ \bibnamefont {MacDowell}},
  \bibinfo {author} {\bibfnamefont {D.~N.}\ \bibnamefont {Sibley}}, \bibinfo
  {author} {\bibfnamefont {B.~D.}\ \bibnamefont {Goddard}},\ and\ \bibinfo
  {author} {\bibfnamefont {S.}~\bibnamefont {Kalliadasis}},\ }\bibfield
  {title} {\bibinfo {title} {The vicinity of an equilibrium three-phase contact
  line using density-functional theory: density profiles normal to the fluid
  interface},\ }\href {https://doi.org/10.1080/00268976.2018.1471223}
  {\bibfield  {journal} {\bibinfo  {journal} {Mol. Phys.}\ }\textbf {\bibinfo
  {volume} {116}},\ \bibinfo {pages} {2239} (\bibinfo {year} {2018})},\ \Eprint
  {https://arxiv.org/abs/https://doi.org/10.1080/00268976.2018.1471223}
  {https://doi.org/10.1080/00268976.2018.1471223} \BibitemShut {NoStop}%
\bibitem [{\citenamefont {Bernardino}\ \emph {et~al.}(2009)\citenamefont
  {Bernardino}, \citenamefont {Parry}, \citenamefont {Rasc{\'o}n},\ and\
  \citenamefont {Romero-Enrique}}]{bernardino09}%
  \BibitemOpen
  \bibfield  {author} {\bibinfo {author} {\bibfnamefont {N.~R.}\ \bibnamefont
  {Bernardino}}, \bibinfo {author} {\bibfnamefont {A.~O.}\ \bibnamefont
  {Parry}}, \bibinfo {author} {\bibfnamefont {C.}~\bibnamefont {Rasc{\'o}n}},\
  and\ \bibinfo {author} {\bibfnamefont {J.~M.}\ \bibnamefont
  {Romero-Enrique}},\ }\bibfield  {title} {\bibinfo {title} {Derivation of a
  nonlocal interfacial model for 3d wetting in an external field},\ }\href@noop
  {} {\bibfield  {journal} {\bibinfo  {journal} {J. Phys.: Condens. Matter}\
  }\textbf {\bibinfo {volume} {21}},\ \bibinfo {pages} {465105} (\bibinfo
  {year} {2009})}\BibitemShut {NoStop}%
\bibitem [{\citenamefont {Alizadeh~Pahlavan}\ \emph {et~al.}(2018)\citenamefont
  {Alizadeh~Pahlavan}, \citenamefont {Cueto-Felgueroso}, \citenamefont {Hosoi},
  \citenamefont {McKinley},\ and\ \citenamefont {Juanes}}]{pahlavan18}%
  \BibitemOpen
  \bibfield  {author} {\bibinfo {author} {\bibfnamefont {A.}~\bibnamefont
  {Alizadeh~Pahlavan}}, \bibinfo {author} {\bibfnamefont {L.}~\bibnamefont
  {Cueto-Felgueroso}}, \bibinfo {author} {\bibfnamefont {A.~E.}\ \bibnamefont
  {Hosoi}}, \bibinfo {author} {\bibfnamefont {G.~H.}\ \bibnamefont
  {McKinley}},\ and\ \bibinfo {author} {\bibfnamefont {R.}~\bibnamefont
  {Juanes}},\ }\bibfield  {title} {\bibinfo {title} {Thin films in partial
  wetting: stability, dewetting and coarsening},\ }\href
  {https://doi.org/10.1017/jfm.2018.255} {\bibfield  {journal} {\bibinfo
  {journal} {J. Fluid Mech.}\ }\textbf {\bibinfo {volume} {845}},\ \bibinfo
  {pages} {642} (\bibinfo {year} {2018})}\BibitemShut {NoStop}%
\bibitem [{\citenamefont {de~Gennes}(1985)}]{degennes85}%
  \BibitemOpen
  \bibfield  {author} {\bibinfo {author} {\bibfnamefont {P.~G.}\ \bibnamefont
  {de~Gennes}},\ }\bibfield  {title} {\bibinfo {title} {Wetting: statics and
  dynamics},\ }\href {https://doi.org/10.1103/RevModPhys.57.827} {\bibfield
  {journal} {\bibinfo  {journal} {Rev. Mod. Phys.}\ }\textbf {\bibinfo {volume}
  {57}},\ \bibinfo {pages} {827} (\bibinfo {year} {1985})}\BibitemShut
  {NoStop}%
\bibitem [{\citenamefont {Davidovitch}\ \emph {et~al.}(2005)\citenamefont
  {Davidovitch}, \citenamefont {Moro},\ and\ \citenamefont
  {Stone}}]{davidovitch05}%
  \BibitemOpen
  \bibfield  {author} {\bibinfo {author} {\bibfnamefont {B.}~\bibnamefont
  {Davidovitch}}, \bibinfo {author} {\bibfnamefont {E.}~\bibnamefont {Moro}},\
  and\ \bibinfo {author} {\bibfnamefont {H.~A.}\ \bibnamefont {Stone}},\
  }\bibfield  {title} {\bibinfo {title} {Spreading of viscous fluid drops on a
  solid substrate assisted by thermal fluctuations},\ }\href
  {https://doi.org/10.1103/PhysRevLett.95.244505} {\bibfield  {journal}
  {\bibinfo  {journal} {Phys. Rev. Lett.}\ }\textbf {\bibinfo {volume} {95}},\
  \bibinfo {pages} {244505} (\bibinfo {year} {2005})}\BibitemShut {NoStop}%
\bibitem [{\citenamefont {Churaev}(1988)}]{churaev88}%
  \BibitemOpen
  \bibfield  {author} {\bibinfo {author} {\bibfnamefont {N.~V.}\ \bibnamefont
  {Churaev}},\ }\bibfield  {title} {\bibinfo {title} {Wetting films and
  wetting},\ }\href {https://doi.org/10.1051/rphysap:01988002306097500}
  {\bibfield  {journal} {\bibinfo  {journal} {Rev. Phys. Appl. (Paris)}\
  }\textbf {\bibinfo {volume} {23}},\ \bibinfo {pages} {975} (\bibinfo {year}
  {1988})}\BibitemShut {NoStop}%
\bibitem [{\citenamefont {de~Gennes}\ \emph {et~al.}(2004)\citenamefont
  {de~Gennes}, \citenamefont {Brochard-Wyart},\ and\ \citenamefont
  {Qu{\'e}r{\'e}}}]{degennes04}%
  \BibitemOpen
  \bibfield  {author} {\bibinfo {author} {\bibfnamefont {P.~G.}\ \bibnamefont
  {de~Gennes}}, \bibinfo {author} {\bibfnamefont {F.}~\bibnamefont
  {Brochard-Wyart}},\ and\ \bibinfo {author} {\bibfnamefont {D.}~\bibnamefont
  {Qu{\'e}r{\'e}}},\ }\href@noop {} {\emph {\bibinfo {title} {Capillarity and
  Wetting Phenomena}}}\ (\bibinfo  {publisher} {Springer},\ \bibinfo {address}
  {New York},\ \bibinfo {year} {2004})\ pp.\ \bibinfo {pages}
  {1--292}\BibitemShut {NoStop}%
\bibitem [{\citenamefont {Starov}\ and\ \citenamefont
  {Velarde}(2009)}]{starov09}%
  \BibitemOpen
  \bibfield  {author} {\bibinfo {author} {\bibfnamefont {V.~M.}\ \bibnamefont
  {Starov}}\ and\ \bibinfo {author} {\bibfnamefont {M.~G.}\ \bibnamefont
  {Velarde}},\ }\bibfield  {title} {\bibinfo {title} {{Surface Forces and
  Wetting Phenomena}},\ }\href@noop {} {\bibfield  {journal} {\bibinfo
  {journal} {J. Phys.: Condens. Matter}\ }\textbf {\bibinfo {volume} {{21}}},\
  \bibinfo {pages} {464121} (\bibinfo {year} {{2009}})}\BibitemShut {NoStop}%
\bibitem [{\citenamefont {Yin}\ \emph {et~al.}(2017)\citenamefont {Yin},
  \citenamefont {Sibley}, \citenamefont {Thiele},\ and\ \citenamefont
  {Archer}}]{yin17}%
  \BibitemOpen
  \bibfield  {author} {\bibinfo {author} {\bibfnamefont {H.}~\bibnamefont
  {Yin}}, \bibinfo {author} {\bibfnamefont {D.~N.}\ \bibnamefont {Sibley}},
  \bibinfo {author} {\bibfnamefont {U.}~\bibnamefont {Thiele}},\ and\ \bibinfo
  {author} {\bibfnamefont {A.~J.}\ \bibnamefont {Archer}},\ }\bibfield  {title}
  {\bibinfo {title} {Films, layers, and droplets: The effect of near-wall fluid
  structure on spreading dynamics},\ }\href
  {https://doi.org/10.1103/PhysRevE.95.023104} {\bibfield  {journal} {\bibinfo
  {journal} {Phys. Rev. E}\ }\textbf {\bibinfo {volume} {95}},\ \bibinfo
  {pages} {023104} (\bibinfo {year} {2017})}\BibitemShut {NoStop}%
\bibitem [{\citenamefont {Dur{\'a}n-Olivencia}\ \emph
  {et~al.}(2019)\citenamefont {Dur{\'a}n-Olivencia}, \citenamefont {Gvalani},
  \citenamefont {Kalliadasis},\ and\ \citenamefont
  {Pavliotis}}]{duran-olivencia19}%
  \BibitemOpen
  \bibfield  {author} {\bibinfo {author} {\bibfnamefont {M.}~\bibnamefont
  {Dur{\'a}n-Olivencia}}, \bibinfo {author} {\bibfnamefont {R.}~\bibnamefont
  {Gvalani}}, \bibinfo {author} {\bibfnamefont {S.}~\bibnamefont
  {Kalliadasis}},\ and\ \bibinfo {author} {\bibfnamefont {G.~A.}\ \bibnamefont
  {Pavliotis}},\ }\bibfield  {title} {\bibinfo {title} {Instability, rupture
  and fluctuations in thin liquid films: Theory and computations.},\
  }\href@noop {} {\bibfield  {journal} {\bibinfo  {journal} {J. Stat. Phys.}\
  }\textbf {\bibinfo {volume} {174}},\ \bibinfo {pages} {579} (\bibinfo {year}
  {2019})}\BibitemShut {NoStop}%
\bibitem [{\citenamefont {Zhang}\ \emph {et~al.}(2020)\citenamefont {Zhang},
  \citenamefont {Sprittles},\ and\ \citenamefont {Lockerby}}]{zhang20}%
  \BibitemOpen
  \bibfield  {author} {\bibinfo {author} {\bibfnamefont {Y.}~\bibnamefont
  {Zhang}}, \bibinfo {author} {\bibfnamefont {J.~E.}\ \bibnamefont
  {Sprittles}},\ and\ \bibinfo {author} {\bibfnamefont {D.~A.}\ \bibnamefont
  {Lockerby}},\ }\bibfield  {title} {\bibinfo {title} {Nanoscale thin-film
  flows with thermal fluctuations and slip},\ }\href
  {https://doi.org/10.1103/PhysRevE.102.053105} {\bibfield  {journal} {\bibinfo
   {journal} {Phys. Rev. E}\ }\textbf {\bibinfo {volume} {102}},\ \bibinfo
  {pages} {053105} (\bibinfo {year} {2020})}\BibitemShut {NoStop}%
\bibitem [{\citenamefont {Saiseau}\ \emph {et~al.}(2022)\citenamefont
  {Saiseau}, \citenamefont {Pedersen}, \citenamefont {Benjana}, \citenamefont
  {Carlson}, \citenamefont {Delabre}, \citenamefont {Salez},\ and\
  \citenamefont {Delville}}]{saiseau22}%
  \BibitemOpen
  \bibfield  {author} {\bibinfo {author} {\bibfnamefont {R.}~\bibnamefont
  {Saiseau}}, \bibinfo {author} {\bibfnamefont {C.}~\bibnamefont {Pedersen}},
  \bibinfo {author} {\bibfnamefont {A.}~\bibnamefont {Benjana}}, \bibinfo
  {author} {\bibfnamefont {A.}~\bibnamefont {Carlson}}, \bibinfo {author}
  {\bibfnamefont {U.}~\bibnamefont {Delabre}}, \bibinfo {author} {\bibfnamefont
  {T.}~\bibnamefont {Salez}},\ and\ \bibinfo {author} {\bibfnamefont {J.-P.}\
  \bibnamefont {Delville}},\ }\bibfield  {title} {\bibinfo {title}
  {Near-critical spreading of droplets.},\ }\href@noop {} {\bibfield  {journal}
  {\bibinfo  {journal} {Nuovo Cimento}\ }\textbf {\bibinfo {volume} {13}},\
  \bibinfo {pages} {7442} (\bibinfo {year} {2022})}\BibitemShut {NoStop}%
\bibitem [{\citenamefont {Kardar}\ \emph {et~al.}(1986)\citenamefont {Kardar},
  \citenamefont {Parisi},\ and\ \citenamefont {Zhang}}]{kardar86}%
  \BibitemOpen
  \bibfield  {author} {\bibinfo {author} {\bibfnamefont {M.}~\bibnamefont
  {Kardar}}, \bibinfo {author} {\bibfnamefont {G.}~\bibnamefont {Parisi}},\
  and\ \bibinfo {author} {\bibfnamefont {Y.-C.}\ \bibnamefont {Zhang}},\
  }\bibfield  {title} {\bibinfo {title} {Dynamic scaling of growing
  interfaces},\ }\href {https://doi.org/10.1103/PhysRevLett.56.889} {\bibfield
  {journal} {\bibinfo  {journal} {Phys. Rev. Lett.}\ }\textbf {\bibinfo
  {volume} {56}},\ \bibinfo {pages} {889} (\bibinfo {year} {1986})}\BibitemShut
  {NoStop}%
\bibitem [{\citenamefont {Wio}(2009)}]{wio09}%
  \BibitemOpen
  \bibfield  {author} {\bibinfo {author} {\bibfnamefont {H.~S.}\ \bibnamefont
  {Wio}},\ }\bibfield  {title} {\bibinfo {title} {Variational formulation for
  the kpz and related kinetic equations},\ }\href
  {https://doi.org/10.1142/S0218127409024505} {\bibfield  {journal} {\bibinfo
  {journal} {Int. J. Bifurcation and Chaos}\ }\textbf {\bibinfo {volume}
  {19}},\ \bibinfo {pages} {2813} (\bibinfo {year} {2009})},\ \Eprint
  {https://arxiv.org/abs/https://doi.org/10.1142/S0218127409024505}
  {https://doi.org/10.1142/S0218127409024505} \BibitemShut {NoStop}%
\bibitem [{\citenamefont {Wio}\ \emph {et~al.}(2022)\citenamefont {Wio},
  \citenamefont {Deza}, \citenamefont {S\'anchez}, \citenamefont
  {Garc{\'i}a-Garc{\'i}a}, \citenamefont {Gallego}, \citenamefont {Revelli},\
  and\ \citenamefont {Deza}}]{wio22}%
  \BibitemOpen
  \bibfield  {author} {\bibinfo {author} {\bibfnamefont {H.}~\bibnamefont
  {Wio}}, \bibinfo {author} {\bibfnamefont {J.}~\bibnamefont {Deza}}, \bibinfo
  {author} {\bibfnamefont {A.}~\bibnamefont {S\'anchez}}, \bibinfo {author}
  {\bibfnamefont {R.}~\bibnamefont {Garc{\'i}a-Garc{\'i}a}}, \bibinfo {author}
  {\bibfnamefont {R.}~\bibnamefont {Gallego}}, \bibinfo {author} {\bibfnamefont
  {J.}~\bibnamefont {Revelli}},\ and\ \bibinfo {author} {\bibfnamefont
  {R.}~\bibnamefont {Deza}},\ }\bibfield  {title} {\bibinfo {title} {The
  nonequilibrium potential today: A short review},\ }\href
  {https://doi.org/https://doi.org/10.1016/j.chaos.2022.112778} {\bibfield
  {journal} {\bibinfo  {journal} {Chaos, Solitons {\&} Fractals}\ }\textbf
  {\bibinfo {volume} {165}},\ \bibinfo {pages} {112778} (\bibinfo {year}
  {2022})}\BibitemShut {NoStop}%
\bibitem [{\citenamefont {Thiele}(2010)}]{thiele10}%
  \BibitemOpen
  \bibfield  {author} {\bibinfo {author} {\bibfnamefont {U.}~\bibnamefont
  {Thiele}},\ }\bibfield  {title} {\bibinfo {title} {Thin film evolution
  equations from (evaporating) dewetting liquid layers to epitaxial growth},\
  }\href {https://doi.org/10.1088/0953-8984/22/8/084019} {\bibfield  {journal}
  {\bibinfo  {journal} {J. Phys.: Condens. Matter}\ }\textbf {\bibinfo {volume}
  {22}},\ \bibinfo {pages} {084019} (\bibinfo {year} {2010})}\BibitemShut
  {NoStop}%
\bibitem [{\citenamefont {Chernov}\ and\ \citenamefont
  {Mikheev}(1988)}]{chernov88}%
  \BibitemOpen
  \bibfield  {author} {\bibinfo {author} {\bibfnamefont {A.~A.}\ \bibnamefont
  {Chernov}}\ and\ \bibinfo {author} {\bibfnamefont {L.~V.}\ \bibnamefont
  {Mikheev}},\ }\bibfield  {title} {\bibinfo {title} {Wetting of solid surfaces
  by a structured simple liquid: Effect of fluctuations},\ }\href
  {https://doi.org/10.1103/PhysRevLett.60.2488} {\bibfield  {journal} {\bibinfo
   {journal} {Phys. Rev. Lett.}\ }\textbf {\bibinfo {volume} {60}},\ \bibinfo
  {pages} {2488} (\bibinfo {year} {1988})}\BibitemShut {NoStop}%
\bibitem [{\citenamefont {Saito}(1980)}]{saito80}%
  \BibitemOpen
  \bibfield  {author} {\bibinfo {author} {\bibfnamefont {Y.}~\bibnamefont
  {Saito}},\ }\bibfield  {title} {\bibinfo {title} {Statics and dynamics of the
  roughening transition: A self-consistent calculation},\ }in\ \href@noop {}
  {\emph {\bibinfo {booktitle} {Ordering in Strongly Fluctuating Condensed
  Matter Systems}}},\ \bibinfo {editor} {edited by\ \bibinfo {editor}
  {\bibfnamefont {T.}~\bibnamefont {Riste}}}\ (\bibinfo  {publisher} {Plenum, New
  York},\ \bibinfo {year} {1980})\ pp.\ \bibinfo {pages} {319--324}\BibitemShut
  {NoStop}%
\bibitem [{\citenamefont {Cuerno}\ and\ \citenamefont {Moro}(2001)}]{moro01b}%
  \BibitemOpen
  \bibfield  {author} {\bibinfo {author} {\bibfnamefont {R.}~\bibnamefont
  {Cuerno}}\ and\ \bibinfo {author} {\bibfnamefont {E.}~\bibnamefont {Moro}},\
  }\bibfield  {title} {\bibinfo {title} {Dynamic renormalization group study of
  a generalized continuum model of crystalline surfaces},\ }\href
  {https://doi.org/10.1103/PhysRevE.65.016110} {\bibfield  {journal} {\bibinfo
  {journal} {Phys. Rev. E}\ }\textbf {\bibinfo {volume} {65}},\ \bibinfo
  {pages} {016110} (\bibinfo {year} {2001})}\BibitemShut {NoStop}%
\end{thebibliography}

%

\clearpage

\onecolumngrid

\setcounter{page}{1}
\setcounter{table}{0}
\setcounter{figure}{0}
\pagenumbering{arabic}

{\centering
{\large Supporting Information for}

{\Large Surface tension of bulky colloids, capillarity under gravity,
	\\  and the microscopic origin of the Kardar-Parisi-Zhang equation
\\ by \\}
{\large  Luis G. MacDowell}

{\normalsize Dpto. de Qu\'{\i}mica F\'{\i}sica, Facultad de Ciencias Qu\'{\i}micas,\\ Universidad Complutense de Madrid, 28040 Madrid, Spain}

}

\vspace*{0.6cm}

This document contains supporting information on the derivation of
results from the main paper. To facilitate cross referencing, this
materials is written as  an appendix section. The
equation numbering and bibliography  follow the original paper, with
equation labels and references
not in this document  referring to those of the original paper.

\vspace*{0.4cm}

\section{Summary of experimental data employed in Figure 1}

In Fig.1, the relation between stiffness coefficients and tilt angle, $\alpha$, is required. Unfortunately, Ref.\cite{thorneywork17} does not provide tabulated data for the stiffness coefficient, and this data could not be obtained upon request from the authors.

In order to map $\tilde{\gamma}$ as a function of $\alpha$, I first obtain the
stiffness as a function of $\beta$ from Figure S5 of Ref.\cite{thorneywork17}
The angle $\beta$, is then mapped into angle $\theta$, according to the
transformation $\beta=15-\theta$, followed by application of sixfold symmetry in
order to guarantee  $\beta$ falls in the range between 0 and 60 degrees. Once a
relation between $\tilde{\gamma}$ and $\theta$ has been made, I use the relation
between $\theta$ and $\alpha$ in table SI of Ref.\cite{thorneywork17} to map
$\tilde{\gamma}$ as a function of $\alpha$. This provides column 7 of Table I
below.

In order to check this result, it is desirable to confirm that indeed, these values
of $\tilde{\gamma}$ are consistent with the independently determined parameters $\langle \hh^2\rangle$ and $L$. 

Unfortunately, table SI from  Ref.\cite{thorneywork17} does not provide the
values of $L$. These can be retrieved from Figure 3(g) of
Ref.\cite{thorneywork17}. Unfortunately, this data is not given as a function of an independent variable, but instead is plotted as a function of
$\sqrt{\tilde{\gamma}/\sin(\alpha)}$. In order to map $L$ as a function of $\alpha$, I assume  $\sqrt{\tilde{\gamma}/\sin(\alpha)}$ changes in inverse proportion to $\alpha$.
This provides the data of the 8th column in Table I below.

\begin{table}[ht]
\begin{tabular}{ccccccccccc}
	\hline
	\multicolumn{1}{c}{$\alpha^{\circ}$} && \multicolumn{1}{c}{$\theta^{\circ}$} && \multicolumn{1}{c}{$\beta^{\circ}$} && \multicolumn{1}{c}{$h_{g\parallel}/\mu m$} & \multicolumn{1}{c}{$h_{g\perp}/\mu m$} & \multicolumn{1}{c}{$\sqrt{\hh^2}/\mu m$} & \multicolumn{1}{c}{$\tilde\gamma\cdot10^{16}/Jm^{-1}$}  & \multicolumn{1}{c}{$L/\mu m$}\\ 
	\hline
	0.560 &&   46.43 && 28.57 &&  $\;$7.0    & 0.068  & $\;\;$8.1  & 2.85 &   $\;\;$9.095 \\
	0.440 &&  45.62  &&  29.38  && $\;$8.9 & 0.068  & $\;\;$9.4  & 1.97 &   $\;\;$8.322 \\
	0.350 &&  30.06  && 44.94  &&11.2  & 0.068  & 10.8 & 1.34 &   $\;\;$7.553 \\
	0.250 &&  58.79  && 16.21  &&15.7  & 0.068  & 12.1 & 1.37 &   $\;\;$9.536 \\
	0.083 && 20.81  &&54.19  &&48.4  & 0.068  & 17.5 & 1.09 & 15.391 \\
	0.067 && 14.31  && $\;\;$0.69  &&55.9  & 0.068  & 22.0   & 0.65 & 15.872 \\
	\hline
\end{tabular}
\caption{Summary of experimental results for the surface fluctuations of hard
discs from Ref.\cite{thorneywork17}. $\alpha^{\circ}$ is the tilt angle of the
colloidal monolayer. $\theta^{\circ}$ is a measure of the orientation of the
hexatic phase with respect to the average interface position. $\beta^{\circ}$	is
a related angle adapted to the hexatic symmetry (see text). $h_{g\parallel}$ and
$h_{g\perp}$ are parallel and perpendicular gravitational heights.
$\langle\hh^2\rangle$ is the mean roughness of the interface. $\tilde{\gamma}$
is the stiffness coefficient and $L=\xi_{\parallel}$ is the parallel correlation
length. All data are from Table S1 in Ref.\cite{thorneywork17}, except for
$\beta$, $\tilde{\gamma}$ and $L$, which are retrieved from analysis of Figures
3(g) and S5 of Ref.\cite{thorneywork17} as explained in the  text.}
\end{table}

Consistency of the data can now be assessed by computing $\tilde\gamma$ from the independently determined parameters $\langle \hh^2\rangle$ and $L$, according to the equation:
\begin{equation}
 \tilde{\gamma}=\frac{1}{2}\frac{k_BT}{\langle\hh^2\rangle} L
\end{equation}
Unfortunately, Ref.\cite{thorneywork17} does not provide data for the temperature, and this could not be obtained upon request from the authors. However, assuming the reasonable value of $T=300$~K, the data for $\langle\hh^2\rangle$ and $L$  of Table S1 provide estimations of $\tilde{\gamma}$ in excellent agreement with the digitalized data of column 7.

\section*{Prove of \Eq{idrjg}}

In order to derive \Eq{idrjg}, we first rewrite \Eq{functional} in condensed notation as:
\begin{equation}\label{eq:functional2}
H[\hh] = \int F(\x;\hh,\hh_{\x}) d{\bf x} 
\end{equation}
with
\begin{equation}\label{eq:funcf}
F(\x;\hh,\hh_{\x})= \int \, \left [ V(z) \rho_{\pi}\left(\frac{z-\hh}{\sqrt{1
		+ 		\hh_{\x}^2}}\right)\right ]d z +    \gamma_0 \, \sqrt{1+\hh_{\x}^2}
- \Delta p\,\hh  
\end{equation}
Here, the subindex $_{\x}$ stands for differentiation with respect to $\x$.

The functional of
\Eq{functional2}  has
an extremal that is given by the Euler-Lagrange equation:
\begin{equation}\label{eq:integrand}
 \frac{\delta H}{\delta \hh(\x)} = \frac{\partial F}{\partial \hh} - \frac{d}{d \x} \left ( \frac{\partial F}{\partial \hh_{\x}} \right)
\end{equation} 

Differentiation of \Eq{funcf} with help of the chain rule
yields:

\begin{equation}\label{eq:euler-lagrange}
\frac{\partial F}{\partial \hh} = -  \frac{\tilde\Pi(\hh,\hh_{\x})}{\sqrt{1 + 	\hh_{\x}^2}} - \Delta p
\end{equation}
and
\begin{equation}
\frac{\partial F}{\partial \hh_{\x}} =  \frac{\Delta \tilde\gamma(\hh,\hh_{\x})  \hh{_{\x}}}{(1 + 	\hh_{\x}^2)^{3/2}}
+ \frac{\gamma_0 \hh_{\x}}{\sqrt{1 + 	\hh_{\x}^2}}
\end{equation}
where: 
\begin{equation}\label{eq:disjnl}
\tilde\Pi(\hh,\hh_{\x}) =  \int \, \left [ V(z) \frac{d\rho_{\pi}}{dz}\left(\frac{z-\hh}{\sqrt{1 + 	\hh_{\x}^2}}\right)\right ]d z
\end{equation}
and
\begin{equation}\label{eq:gammanl}
\Delta \tilde\gamma(\hh,\hh_{\x}) =  -\int \, \left [ (z-\hh)V(z) \frac{d\rho_{\pi}}{dz}\left(\frac{z-\hh}{\sqrt{1 + 	\hh_{\x}^2}}\right)\right ]d z
\end{equation}

Replacing these results into \Eq{euler-lagrange}, gives
the following stationarity condition for $\hh(\x)$:
\begin{equation}\label{eq:extremalagain}
-  \frac{\tilde\Pi(\hh,\hh_{\x})}{\sqrt{1 + 	\hh_{\x}^2}} - \Delta p = \frac{d}{dx}\left ( \frac{\gamma_0 \hh_{\x}}{\sqrt{1 + 	\hh_{\x}^2}} + \frac{\Delta \tilde\gamma(\hh,\hh_{\x})  \hh{_{\x}}}{(1 + 	\hh_{\x}^2)^{3/2}}
 \right )
\end{equation}

Notice that, whereas both $\tilde{\Pi}$ and $\Delta\tilde\gamma$ stem from the external field, the former plays the role of a disjoining pressure, while the latter effectively appears as a correction to the surface tension.  The explicit dependence of these functions on $\hh_{\x}$ is a consequence of the non-locality of  the free energy functional,  \Eq{functional2}  with respect to $\hh(\x)$ (i.e. the non-local dependece of $\tilde{\Pi}$ and $\Delta\tilde\gamma$ on $\hh$ can be cast approximately in terms of local functions of  $\hh(\x)$ {\em and} $\hh_{\x}(\x)$).

The integrals of \Eq{disjnl} and \Eq{gammanl} cannot be evaluated in closed form without further assumptions. However, we notice that the derivative of the density profile can be considered  to leading order as a  sharp symmetrical distribution centered at $z=h$. For external fields varying smoothly in the scale of one correlation length, as is usually the case, we can therefore  expand $V(z)$ in the integrand about $z=h$. To leading order in the expansion, this yields:
\begin{equation}
  \tilde\Pi(\hh,\hh_{\x}) =  -V(\hh)\int \, \left [ \frac{d\rho_{\pi}}{d\hh}\left(\frac{z-\hh}{\sqrt{1 + 	\hh_{\x}^2}}\right)\right ]d z
\end{equation}
A simple change of variables then leads to the convenient approximation:
\begin{equation}\label{eq:disjfactorized}
 \tilde\Pi(\hh,\hh_{\x})\approx \Pi(\hh) \sqrt{1 + 	\hh_{\x}^2}
\end{equation}
where $\Pi(\hh)$ is the disjoining pressure of a planar interface.

\rechange{
In order to evaluate $\Delta \tilde\gamma$, we notice that, to a good
approximation $t\frac{d\rho_{\pi}}{dz}(t) = -\xi^2 \frac{d^2\rho_{\pi}}{dt
dz}(t)$,  where $t$ is an arbitrary variable, and
$\xi$ is a measure of the interfacial width.\cite{macdowell13,benet14b,macdowell14,macdowell17}
By taking this into account, we can write:
\begin{equation}
\Delta \tilde\gamma(\hh,\hh_{\x}) = - \left( 1 +    \hh_{\x}^2 \right) \int \,
V(z)  \frac{d}{d h} \frac{d\rho_{\pi}}{dz}\left(\frac{z-\hh}{\sqrt{1 + 	\hh_{\x}^2}} \right ) d z
\end{equation}
A simple rearrangement, followed by comparison with \Eq{disjnl}, leads to the
convenient result:
\begin{equation}\label{eq:gammafactorized}
 \Delta \tilde\gamma(\hh,\hh_{\x}) \approx - \left( 1 +    \hh_{\x}^2 \right)\xi^2 \frac{d \tilde \Pi}{d\hh}
\end{equation}
}

\rechange{
Finally, replacing \Eq{disjfactorized} and \Eq{gammafactorized} into
\Eq{extremalagain}, leads to:
	\begin{equation}\label{eq:correct}
	   \Pi(\hh) + \Delta p = -\frac{d}{d\x}\left (  \frac{\gamma_0
	   \hh_{\x}}{\sqrt{1 +    \hh_{\x}^2}}  + \Delta\gamma(\hh)  \hh_{\x}   \right )
	\end{equation}
   } which, in the limit of small gradient leads to the sough result.

\Eq{correct} corrects the results from a preliminary version of this article
(arXiv:2302.01959). The result also shows that the linearized form of the
equilibrium condition published in Ref.\cite{benet14b} is innacurate. In that
paper, \Eq{extremalagain} was linearized, and it was assumed that the non-local
functionals $\tilde\Pi(\hh,\hh_{\x})$ and $\Delta \tilde\gamma(\hh,\hh_{\x})$
could be approximated by their local forms for the flat profile, i.e. $\Pi(\hh)$
and $\Delta\gamma(\hh)$, respectively. This appears to be incorrect in view of
the above. 

\section{Prove of \Eq{1stint}}

To obtain \Eq{1stint}, consider the one dimensional film profile, $h(x)$ of a cylindrical drop or liquid wedge along the $x$ direction. For this problem, the equilibrium condition, \Eq{idrjg} simplifies to:

\rechange{
\begin{equation}
 \Pi(\hh) + \Delta p = -\frac{d}{dx}\left ( \gamma(\hh) \hh_{x} \right )
\end{equation}
}
where $\hh_x$ denotes derivation with respect to $x$.

Multiplying this result by $d\hh$, the equilibrium condition may be cast as:
\rechange{
\begin{equation}
\left ( \Pi(\hh) + \Delta p \right ) d\hh = -\hh_{x} d\left ( \gamma(\hh) \hh_{x} \right )
\end{equation}
} 
The right hand side of this equation obeys:
\rechange{
\begin{equation}
 h_x d\left ( \gamma(\hh) \hh_{x}\right ) =  d\left ( \gamma(\hh) \hh_{x}^2 \right ) - 
 \frac{1}{2}\gamma(\hh) d \hh_{x}^2 
\end{equation}
}
so that one can write exactly:
\begin{equation}
d\left ( \gamma(\hh) \hh_{x}^2 \right ) + \left ( \Pi(\hh) + \Delta p \right ) d\hh = \frac{1}{2} \gamma(\hh)  d \hh_{x}^2 
\end{equation}
This result is now integrated from $\hh(x)=\hh_e$ at $x\to-\infty$, where $\hh_x(x)=0$, to $\hh(x)$ at arbitrary $x$, leading to:

\begin{equation}\label{cosica}
 \gamma(\hh) \hh_{x}^2 -( \omega(\hh) - \omega(\hh_e)) = \frac{1}{2}\int_{\hh_e}^{\hh} \gamma(\hh) \frac{d \hh_{x}^2}{d \hh} d\hh
\end{equation}
where $\omega(\hh) = g(\hh) - \Delta p \hh$. 

This result remains also a complex
integro-differential equation, but is now amenable to an approximate solution upon succesive iteration.

To see this, first solve under the assumption that $\gamma(\hh)$ is a constant equal to $\gamma_0$. This leads right away to:
\begin{equation}
  \hh_x^2 = 2 \frac{\omega(\hh) - \omega(\hh_e)}{\gamma_0}
\end{equation}
which corresponds to the exact first integral of the  Derjaguin or augmented
Young-Laplace equation (c.f. Ref.\cite{churaev88,degennes04,starov09}). 

Now, replacing this result back into the right hand side of Eq.(27), followed by a change of variables in the integrand of the right hand side, yields:
\begin{equation}
 \gamma(\hh) \hh_{x}^2 -( g(\hh) - g(\hh_e)) = \int_{\hh_e}^{\hh} dg +\int_{\hh_e}^{\hh} \frac{1}{2}\frac{\xi^2}{\gamma_0} g' d g'
\end{equation}
where it is assumed the system is exactly at coexistence, such that $\Delta p=0$ and
$\omega(\hh)=g(\hh)$.

This equation leads readily to \Eq{1stint} upon integration. The film profile is then obtained by numerical quadrature.

A relation between the contact angle, $\theta$ and $g(\hh_e)$ may be obtained by noticing that for the choice $\Delta p = 0$, the droplet has zero curvature. Therefore,  
it is acknowledged that as $x\to\infty$, $\hh(x)\to\infty$, and $\hh_x(x)\to\tan\theta$. Applying this condition in \Eq{1stint}, one readily finds that $\frac{1}{2} \gamma_0 \tan^2\theta=-g(\hh_e)$.
In practice, to the order of small gradients that this result applies, $\tan(\theta)=\theta$, so the relation simplifies to $\theta^2 = -\frac{2g(\hh_e)}{\gamma_0}$.

\section{Model interface potential}

The results of Fig.\ref{fig:droplet} are obtained for a model interface potential with a short range contribution and a long range tail favoring wetting:
\begin{equation}
  g(\hh) = C_2 e^{-2\kappa \hh} - C_1 e^{-\kappa \hh} - \frac{A}{12\pi\hh^2}
\end{equation}
where $\kappa$ is the inverse correlation length, $C_i$ are positive constants, and $A$ is the Hamaker constant. In the explicit calculations, these parameters are set to
$C_2/\gamma_0=1$, $C_1/\gamma_0=24$, $\frac{A\kappa^2}{12\pi\gamma_0} = -12$. This leads to a minimum at $\kappa \hh_e\approx 1.88$, with $g(\hh_e)/\gamma_0=-0.24$, and a contact angle of $\theta \approx 40$ degrees.
For the calculation of $\Delta \gamma(\hh)$, a value of the interfacial width of $\xi=\sqrt{3}\kappa^{-1}$ is assumed (based on comparison of \Eq{general} with results for an exact model.\cite{macdowell17}).

\section{Prove of \Eq{kpzm}}

From the proof of \Eq{idrjg}, the first functional derivative of \Eq{functional} is:
\begin{equation}
\frac{\delta H}{\delta \hh(\x)} = -  \Pi(\hh) - \Delta p -  \frac{d}{dx}\left ( \frac{\gamma(h) \hh_{\x}}{\sqrt{1 + 	\hh_{\x}^2}} \right )
\end{equation}
To quadratic order in $\hh_{\x}$, this leads to:
\begin{equation}
\frac{\delta H}{\delta \hh(\x)} = -  \Pi(\hh) - \Delta p - 
\gamma(\hh)\hh_{\x\x} - \gamma'(\hh) \hh_{\x}^2
\end{equation}
Replacing this result in the equation for non-conserved dynamics under the assumption of phase coexistence ($\Delta p=0$):
\begin{equation}
\frac{\partial \hh}{\partial t} = -\frac{\delta H}{\delta \hh(\x)}
\end{equation}
leads right away to \Eq{kpzm}.

\end{document}